\documentclass[]{svjour3}
\usepackage{graphicx}
\usepackage{epstopdf}
\usepackage{amsmath,amssymb,amsfonts}
\usepackage{mathtools}
\usepackage{multirow}
\usepackage{verbatim}
\usepackage{url}
\usepackage{comment}
\usepackage{mathtools}
\usepackage{epsf,pgf,graphicx}
\usepackage{color}
\usepackage{tikz,pgf}
\usepackage{diagbox}
\usepackage{makecell}
\usepackage{color}
\usepackage{braket}
\usepackage{mathtools}
\usepackage{xcolor}
\usepackage{subfig}
\usepackage{hyperref}
\usepackage{cite}

\usepackage{tabstackengine}
\stackMath
\usepackage{array}
\newcolumntype{M}[1]{>{\hbox to #1\bgroup\hss$}l<{$\egroup}}

\makeatletter
\newcommand\@brcolwidth{0.67em}

\def\@brarray[#1]{\array{r*\c@MaxMatrixCols {M{#1}}}}
\makeatother

\title{On Obtaining New MUBs by Finding Points on Complete Intersection Varieties over $\mathbb{R}$}
\titlerunning{Obtaining New MUBs by Finding Points on Complete Intersection Varieties}
\author{Arindam Banerjee \and Kanoy Kumar Das \and Ajeet Kumar \and Rakesh Kumar \and Subhamoy Maitra}
\institute{
A. Banerjee \at
Indian Institute of Technology, Kharagpur, India,
\email{123.arindam@maths.iitkgp.ac.in}
\and
K. K. Das \at
Chennai Mathematical Institute, Chennai, India,
\email{kanoydas0296@gmail.com}
\and
Ajeet Kumar \at
Indian Statistical Institute, Kolkata, India,
\email{ajeetk52@gmail.com}
\and
R. Kumar \at
Indian Statistical Institute, Kolkata, India,
\email{rkmath1729@gmail.com}
\and
S. Maitra \at
Indian Statistical Institute, Kolkata, India,
\email{subho@isical.ac.in}
}
\begin{document}
\maketitle
\begin{abstract}
Mutually Unbiased Bases (MUBs) are closely connected with quantum physics and the structure has a rich mathematical background. We provide equivalent criteria for extending a set of MUBs for $C^n$ by studying real points of a certain affine algebraic variety. This variety comes from the relations that determine the extendability of a system of MUBs. Finally, we show that some part of this variety gives rise to complete intersection domains. Further, we show that there is a one-to-one correspondence between MUBs and the maximal commuting classes (bases) of orthogonal normal matrices in $\mathcal M_n({\mathbb{C}})$. It means that for $m$ MUBs in $C^n$, there are $m$ commuting classes each consisting $n$ commuting orthogonal normal matrices and the existence of maximal commuting basis for $\mathcal M_n({\mathbb{C}})$ ensures the complete set of MUBs in $\mathcal M_n({\mathbb{C}})$.
\end{abstract}

\section{Introduction} 
Mutually Unbiased Bases (MUBs) have been well-studied in quantum information theory and Mathematical science. Its notion was initiated by Schwinger \cite{schwinger1960unitary}, a long back in sixties. The motivation was related to unitary operator bases and their connection to maximal incompatibility. Further Ivanovic \cite{ivonovic1981geometrical} explored the application of MUBs in the context of quantum state determination.  

In the Hilbert space $C^{n}$, two orthonormal bases $A \equiv\{\ket{a_{1}}, \ket{a_{2}}, \ldots, \ket{a_{n}} \}$ and $B\equiv \{\ket{b_{1}}, \ket{b_{2}}, \ldots, \ket{b_{n}}\}$ are mutually unbiased if,
 \begin{equation*}
 | \braket{a_{i}\vert b_{j}} |= \frac{1}{\sqrt{n}},\ \forall i,j=1, 2, \ldots, n.  
 \end{equation*}
The interesting feature of a pair of MUBs is that the overlaps between any two vectors from the two different bases are uniform. That means, a measurement over one basis leaves one completely uncertain as to the outcome of a measurement over a basis unbiased concerning the first. This unbiased-ness property is crucial in quantum cryptography and quantum key distribution protocols \cite{BB84, planat2006survey}.
 
The maximum possible number of Mutually Unbiased Bases (MUBs) in a complex Hilbert space of dimension \( n \) is  \( n+1 \). If \( n+1 \) MUBs exist, then it's possible to optimally determine the state of a quantum system described by a Hilbert space of dimension \( n \) \cite{wootters1989optimal}.
 
The construction of a complete set of MUBs is well understood when the dimension \( n \) is a prime power, i.e., \( n = p^k \), where \( n \) is a prime number and \( k \in \mathbb{N} \) \cite{wootters1989optimal, Bandyopadhyay}. However, the exact number of MUBs is not known in other dimensions, not even for $n=6$. At least three MUBs exist in any dimension $n$. Zauner \cite{zauner2011quantum} has conjectured that there cannot be more than three MUBs in dimension $6$. Numerical evidences appear to support Zauner's conjecture \cite{brierley2008maximal,grassl2004sic}. Although computer-aided analytical techniques like Gr\"obner bases and semidefinite programming have been used to tackle this issue, no conclusive answer has yet been found due to computational power constraints
\cite{brierley2010mutually}.

Now let us briefly introduce the importance of this topic related to theoretical physics. Many fundamental concepts, like quantum entanglement, quantum coherence, etc., in quantum mechanics are connected with MUBs. For example, coherence is basis-dependent. A state may be coherent in one basis but not in another, so the performance of quantum coherence under different bases might be interesting. Baumgratz and others gave an idea for a coherence measure \cite{baumgratz2014quantifying}. Geometric coherence, as one of the measures of coherence, can effectively quantify quantum coherence. In \cite{sun2024applications}, authors connects measures of quantum coherence with MUBs. Further, the MUBs are crucial in constructing maximally entangled states, and entanglement witnesses is one of the tool to characterize quantum entanglement \cite{terhal2000bell}. In \cite{li2019mutually, wang2021constructing}, the study of entanglement witness has been presented in the context of MUBs. Moreover, towards application domain, MUBs are important in designing quantum secret sharing (QSS) schemes. QSS is a protocol where a quantum secret (usually a qubit or a qudit) is distributed among several parties such that only specific subsets of them can reconstruct the secret. In \cite{tavakoli2015secret}, authors provided a secret sharing scheme for $n$-dimensional quantum system where $n$ is an odd prime, and in \cite{hao2019new}, discussion has been done for $ n^2$-dimensional quantum system with $n$ is an odd prime. Finally, it is well known that MUBs are directly related to Quantum Key Distribution (QKD)~\cite{BB84}, in particular for the six-state scenario~\cite{BR98}. This area has deep impact in the domain of quantum cryptography.

From the combinatorial point of view, there exist a series of deep literature too. In the space $C^{n}$, each MUB can be considered as a unitary $n\times n$ matrix. Two such MUBs can be associated with two unitary matrices, among which one can always be transformed into the identity matrix $\emph{I}$ and another into Hadamard matrix (absolute value of every matrix elements are $\frac{1}{\sqrt{n}}$) through an unitary transformation. This representation of MUBs establishes a link between MUBs and complex Hadamard matrices \cite{tadej2006concise, Weigert}.

The challenge in achieving a deeper comprehension of mutually unbiased bases arises from the unavailability of an appropriate mathematical tool to investigate the problem. However, research on MUBs reveals deep connections between the fundamentals of quantum theory and many areas of mathematics, such as group theory, Galois rings, Graph theory,  finite fields, combinatorial designs, and projective geometry \cite{finitepro,bengtsson,kibler, boykin,paterekdakic,paterekpawlowski,kolountzakis}.

In this article, we study various problems related to Mutually Unbiased Bases (MUBs) through certain results of algebraic geometry. We approach the problem in the following way. Once we have $k$ given MUBs in some given dimension $n$, the problem of finding the $(k+1)$-th MUB is essentially the problem of finding solutions to a system of polynomial equations in $2n^2$ real variables and having degree at most two. This is obtained by taking the real and imaginary parts of each complex variable separately and writing down different equations for the real and imaginary parts. This has the benefit of converting the problem of finding a new MUB into a problem of algebraic geometry (taking a complex variable has the disadvantage that in that case conjugation operation, an inherently non-polynomial operation, will remain in play). Under this set up we show that the existence or non-existence of some MUB problems becomes problems of finding real points on some algebraic varieties.
  
\subsection{Organization and Contribution}
In the next section, we present certain preliminaries and background related to this work. The contributory results are in Sections~\ref{sec3} and~\ref{sec4}. 

First, we set up the problem of finding a new MUB as a problem of algebraic geometry in Section~\ref{sec31}. The ideals for dimension 2 are explained in Section~\ref{sec32}. Technically, our following result is important in this direction.
\ \\
\ \\
{\bf Theorem} [Theorem \ref{real points}].
Let $A_1, A_2,\ldots , A_k$ be a system of $k$ mutually unbiased  bases. Then this system can be extended to a system of $k+1$ MUBs if and only if the algebraic variety $Z(M_{k+1}^{\{n\}})$ of the ideal $M_{k+1}^{\{n\}}=I^{\{n\}}+ \sum_{i=1}^{k} J_{l,k+1}^{\{n\}}\subseteq \mathbb{R}[x_{ij}, y_{ij}: 1\leq i,j\leq n]$ has at least one point in $\mathbb R^{2n^2}$.

We next analyze the case through an example to recover the complete set of MUBs for dimension two and recover some known results. Then we show that if we start with the identity matrix, then the system of equations to generate the next MUB contains a part consisting of equations of spheres and some homogeneous equations. The result explains that the ideal generated by the sphere part is a complete intersection prime ideal. This is presented in Section~\ref{sec33}.

Finally, in~\ref{sec4}, we show a connection between MUBs and the maximal commuting classes (bases) of orthogonal normal matrices in $C^{n}$ instead of commuting bases containing orthogonal unitary matrices \cite{Bandyopadhyay}. This observation of considering normal matrices over unitary matrices provides more flexibility. This connection reveals the necessary condition for the existence of MUB in any dimension. Also, we present an example for dimension four, where there are five commuting classes, and each consists of four orthogonal normal matrices.
\ \\
\ \\
{\bf Theorem} [Theorem \ref{sphere}].
	Let  $A_1,A_2, \ldots , A_{m-1}$ be a system of MUBs, and assume that $A_1=I_n$, the identity matrix. Then  $J_{1,m}^{\{n\}}$, the ideal generated by the sphere equations, is a complete intersection prime ideal.

The significance of this result is that if we pass to the quotient ring obtained by modding out the polynomial ring by these equations the dimension reduction is maximum, i.e., same as the number of equations. The rings of this type are very much similar to polynomial rings. This allows us potentially to bring down the MUB finding problem to the study of a ring of potentially much lower dimension that inherits many nice properties of the larger polynomial ring.

\section{Preliminaries}
In this section, we collect all the necessary definitions from commutative algebra and algebraic geometry that are necessary for studying the ideals arising from MUBs. For this, we closely follow the standard text \cite{Matsumura1987}. The definitions we present here may be stated in a more general setup, but for our purpose, we will restrict ourselves to the case where the ring $R$ is a polynomial ring over a field $\mathbb K$.

\begin{definition}
Let $R$ be a ring and $a\neq 0$ be an element of $R$. Then $a$ is said to be a non-zero divisor in $R$ if for any $b\in R$, $ab=0$ implies that $b=0$; otherwise $a\in R$ is said to be a zero divisor. If a ring $R$ has no non-zero zero divisors, then $R$ is called an Integral Domain.
\end{definition}

\begin{definition}
Let $R$ be a ring. An ideal $I\subseteq R$ is said to be a prime ideal if one of the following conditions is satisfied:
\begin{enumerate}
\item $R/I$ is an integral domain.
\item For any $a, b\in R$, if $ab\in I$, then either $a\in I$, or $b\in I$. 
\end{enumerate}
We shall denote $Spec(R)$ to be the set of all prime ideals of the ring $R$.
\end{definition}
Let $R$ be a ring and consider a chain of prime ideals \[P_0 \subsetneq P_1 \subsetneq \cdots \subsetneq P_n.\] The integer $n$ is called the \textit{length} of the above chain of prime ideals. We define the \textit{Krull dimension} of the ring $R$ to be the supremum of the integers $n$ so that there is a chain of prime ideals of length $n$ in the ring $R$. 

Let $P$ be a prime ideal of $R$. Then the \textit{height} of $P$, denoted by $ht(P)$, defined to be the supremum of the length of the chains of prime ideals which ends at $P$, i.e., \[P_0 \subsetneq P_1 \subsetneq \cdots \subsetneq P_n = P.\]
One can attach various numerical values to an ideal to measure the complexity of that ideal, and \textit{height of an ideal} is certainly one of them.
    \begin{definition}
        Let $R$ be a ring and $I\subseteq R$ an ideal. Then the \textit{height} of the ideal $I$, denoted by $ht(I)$, is defined by \[ht(I)=min\{ht(P)|I\subseteq P \text{ and } P\in Spec(R)\}.\]
    \end{definition}
    \begin{definition}
        A sequence $\mathbf{x} = x_1, \ldots, x_n$ of elements of $R$ is called an $R$-regular sequence, or simply, a \textit{regular sequence}, if the following conditions are satisfied:
        \begin{enumerate}
            \item $x_i$ is a non-zero divisor in $R/(x_1,\ldots ,x_{i-1})$ for all $1\leq i\leq n$,
            \item $R/\mathbf{x}R\neq 0$.
        \end{enumerate}
    \end{definition}
The notion of complete intersection comes from algebraic geometry. It essentially means that the defining ideal of an affine algebraic variety is generated by the least possible number of elements when compared to the co-dimension of the affine variety, and the affine variety is simply the intersection of hypersurfaces coming from the generators of the defining ideal.
\begin{definition}
Let $R$ be a ring and $I\subseteq R$ an ideal. We say that $I$ is a complete intersection if $I$ is generated by a regular sequence.
\end{definition}
Complete intersection ideals enjoy a variety of nice properties. The following proposition tells us that in the right setup, complete intersection ideals give rise to Cohen-Macaulay rings.
\begin{proposition}
Let $(R,\mathbf m)$ be a Cohen–Macaulay local ring and $I\subseteq R$ an ideal. If $I$ is a complete intersection, then $R/I$ is Cohen–Macaulay.
\end{proposition}

One of the most fundamental results in the area of commutative algebra and algebraic geometry is Hilbert's Nullstellensatz. We shall use this theorem to translate the problem of extending a system of MUBs to a problem of studying some affine algebraic variety. Let $\mathbb K$ be a field and $\overline{\mathbb K}$ its algebraic closure. Suppose that $\Phi \subseteq \mathbb K[x_1, \ldots, x_n]$ is a subset. An $n$-tuple $\alpha=(a_1, \ldots, a_n)$ of elements $a_i\in \overline{\mathbb K}$ is an \textit{algebraic zero} of $\Phi$ if it satisfies $f(\alpha)=0$ for all $f\in \Phi$. Given a subset $\Phi\subseteq \mathbb K[x_1,\ldots , x_n]$, we shall denote $Z(\Phi)$ to be the set of all algebraic zeros of the set $\Phi$.

    \begin{theorem}\cite[Theorem 5.4]{Matsumura1987}
        \begin{enumerate}
            \item If $\Phi$ is a subset of $\mathbb K[x_1, \ldots, x_n]$ which does not have any algebraic zeros then the ideal generated by $\Phi$ contains $1$.
            \item Given a subset $\Phi \subseteq \mathbb K[x_1, \ldots, x_n]$ and an element $f \in \mathbb K[x_1, \ldots, x_n]$, suppose that $f$ vanishes at every algebraic zero of $\Phi$. Then there exists $t\geq 1$, $g_i\in \mathbb K[x_1, \ldots, x_n]$ and $h_i\in \Phi$ such that $f^t=\sum g_ih_i$. 
        \end{enumerate}
    \end{theorem}
\noindent
   
We now recall the notion of Gr\"obner bases which is a very useful tool in proving various results in commutative algebra. We refer the reader to the standard texts \cite{HerzogHibi1993,HerzogHibiOshugi2018} for a quick review of the theory of Gr\"obner bases. For the rest of the article, we shall denote $S$ to be a polynomial ring over the field $\mathbb{R}$, and $Mon(S)$ will denote the set of all monomials in the polynomial ring $S$. Recall that a \textit{total order} on a set $P$ is a partial order $\leq $ on $P$ with the following extra condition: for any two elements $x, y\in P$, one has either $x\leq y$ or $y\leq x$. A \textit{monomial order} on $S$ is a partial order $<$ on $Mon(S)$ such that the following conditions are satisfied:
	\begin{enumerate}
        \item $<$ is a total order;
		\item for any $ u \in Mon(S)$, and $u\neq 1$ one has $1<u$;
		\item for any $u, v \in Mon(S)$ with $u<v$, one has $uw<vw$ for all $w \in Mon(S)$.
	\end{enumerate}
	A few examples of monomial orders include lexicographical order, reverse lexicographical order, etc. From now on, we will fix a monomial order $<$ on $S$, and all the definitions and results mentioned afterward will use this monomial order only. Let $f=\sum_{v\in Mon(S)}a_v v$ be a non-zero polynomial in $S$ with $a_v\in \mathbb{R}$. The \textit{support} of the polynomial $f$, denoted by $Supp(f)$, is defined to be the collection of all the monomials $v$ in $f=\sum_{v\in Mon(S)}a_v v$, where $a_v\neq 0$. The \textit{initial monomial} of $f$ with respect to the monomial order $<$, denoted by $in_<(f)$,  is the biggest monomial with respect to the monomial order $<$ among all the monomials in $Supp(f)$.
\begin{definition}
Let $I\neq 0$ be an ideal of $S$. The initial ideal of $I$ with respect to the monomial order $<$, denoted by $in_<(I)$, is defined by \[in_<(I):= (\{in_<(f):f\in I\}).\]
Note that, for any given ideal $I\subseteq S$, the initial ideal $in_<(I)$ is always an ideal generated by monomials.
\end{definition}
\begin{definition}
Let $I\neq 0$ be an ideal of $S$. A set of nonzero polynomials $\{f_1, f_2, \ldots$, $f_s\} \subseteq I$, is said to be a Gr\"obner basis of $I$ with respect to $<$ if \[in_<(I)=(in_<(f_1), in_<(f_2), \ldots, in_<(f_s)).\] 
\end{definition}
A Gr\"obner basis of $I$ with respect to $<$ always exists, but need not be unique. Given a monomial order $<$, there is a general notion of division algorithm for a given set of polynomials. We refer the reader to \cite[Thm 2.2.1]{HerzogHibi1993} for a detailed explanation of this notion. Now we recall a very important theorem in the theory of Gr\"obner bases, known as the Buchburger's Criterion, which gives a necessary and sufficient condition for a set of polynomials to be a Gr\"obner basis of an ideal. However, before that, we need some definitions. For any two monomials $u,v\in S$, we shall denote $lcm(u,v)$ to be the least common multiple of $u$ and $v$.
\begin{definition}
Let $f, g\in S$ be two non-zero polynomials. Let $c_f, c_g$ denote the coefficient of $in_<(f), in_<(g)$ in $f$ and  $g$ respectively. The S-polynomial of $f$ and $g$ is the polynomial given by
		\[S(f,g)=\frac{lcm(in_<(f),in_<(g))}{c_fin_<(f)}f-\frac{lcm(in_<(f),in_<(g))}{c_gin_<(g)}g.\]
\end{definition}
We say that a polynomial $f$ reduces to $0$ with respect to the polynomials $f_1, f_2, \ldots$, $f_m$, if in the division algorithm, there is a standard expression of $f$ with respect to $f_1, f_2, \ldots$, $f_m$, such that the remainder will be the zero polynomial. The following lemma will be crucial in some of the results that we prove in later sections.
	
\begin{lemma}\cite[Lemma 2.3.1]{HerzogHibi1993}\label{lcm}
Let $f, g\neq 0$ be two polynomials and suppose that $in_<(f)$ and $in_<(g)$ are relatively prime to each other, i.e. $lcm(in_<(f), in_<(g)) = in_<(f)in_<(g)$. Then $S(f,g)$ reduces to $0$ with respect to $f, g$.
\end{lemma}
The following theorem is central to the theory of Gr\"obner bases which gives a necessary and sufficient condition for a system of generators of an ideal to be a Gr\"obner basis of the ideal.
\begin{theorem}\cite[Buchberger's Criterion]{HerzogHibi1993}\label{BBC}
Let $I\neq 0$ be an ideal of $S$ and $G=\{f_1, f_2$, $\ldots, f_m\}$ a system of generators of $I$. Then $G$ is a Gr\"obner basis of $I$ if and only if for all $i \neq j$, $S(f_i, f_j)$ reduces to $0$ with respect to  $f_1, f_2, \ldots, f_m$.
\end{theorem}

\section{Studying Real Points on Intersection Varieties}
\label{sec3}
In this section, we will formally establish the problem of extending a system of MUB using the tools of algebraic geometry. Additionally, we will explore specific cases for the dimension 2. Subsequently, we will demonstrate how a particular ideal, arising from the investigation of this problem, exhibits some desirable properties in this regard.

\subsection{Defining ideals of MUB}
\label{sec31}
In this subsection, we associate the problem of extending a system of MUBs with the problem of finding points in an affine algebraic variety.  We begin this section with the definition of MUBs.
\begin{definition}
\label{def_MUB}
Two orthonormal bases $A \equiv\{\ket{a_{1}}, \ket{a_{2}}, \ldots, \ket{a_{n}} \}$ and $B \equiv \{\ket{b_{1}}$, $\ket{b_{2}}, \ldots, \ket{b_{n}}\}$ in an $n$ dimensional Hilbert spaces are mutually unbiased if
$$|\braket{a_{i}\vert b_{j}}|= \frac{1}{\sqrt{n}},\ \forall i,j=1, 2, \ldots, n.$$
\end{definition}
If we consider the vector space $\mathbb C^n$, the condition given in the above definition becomes $\|{\mathbf{a}_i}\cdot \overline{{\mathbf{b}_j}}\|=\frac{1}{\sqrt{n}}$ for any two orthonormal bases $\{\mathbf{a}_i: 1\leq i\leq n\}$, and $\{\mathbf{b}_i: 1\leq i\leq n\}$ of $\mathbb C^n$, where for a vector $\mathbf{v}=(v_1, \ldots, v_n)\in \mathbb C^n$, we denote $\overline{\mathbf{v}}:=(\overline{v_1}, \ldots, \overline{v_n})^t$, the vector consisting of the complex conjugates transpose of the row vector $\mathbf{v}$.
	
Let $A=\{\mathbf{a_j} = (a_{1j}, a_{2j}, \ldots, a_{nj}): 1\leq j\leq n\}$ be a set of vectors in the vector space $\mathbb C^n$. For $A$ to be a part of a MUB, $A$ need to be orthonormal. More precisely, we must have $\mathbf{a_j}\cdot \overline{\mathbf{a_k}}=0$ for any $1\leq j\neq k\leq n$. We formulate this criterion in terms of algebraic equations as follows.
	
Let  $z_{ij}, 1 \leq i, j \leq n$ be a set of complex variables. We write $\mathbf{z_j} = (z_{1j}, z_{2j}, \ldots$, $z_{nj})$, where $1 \leq j \leq n$. Any orthogonal set of $n$-vectors of $\mathbb{C}^n$ is a non-zero complex solution of the system of equations $\mathbf{z_j}\cdot \overline{\mathbf{z_k}}=0, 1\leq j\neq k\leq n$. Note that the equations $\mathbf{z_j}\cdot \overline{\mathbf{z_k}}=0$ are not necessarily polynomial equations.
	
To overcome this problem, we reduce this system of complex equations to a system of real equations by making the substitution $z_{ij} = x_{ij} + iy_{ij}$, where $x_{ij}, y_{ij}$ are real variables for all $1\leq i, j \leq n$. Then the above orthogonality relations become 
\begin{align*}
		\mathbf{z_j}\cdot \overline{\mathbf{z_k}}
		&=\sum_{i=1}^{n} z_{ij}\cdot \overline{z_{ik}} 
		=\sum_{i=1}^{n} (x_{ij}+iy_{ij})(x_{ik}-iy_{ik})\\ 
		&= \sum_{i=1}^{n} \{(x_{ij}x_{ik}+y_{ij}y_{ik}) + i(x_{ik}y_{ij}-x_{ij}y_{ik})\}\\
		&=\sum_{i=1}^{n} (x_{ij}x_{ik}+y_{ij}y_{ik}) + i\sum_{i=1}^{n} (x_{ik}y_{ij}-x_{ij}y_{ik}).
\end{align*}
Now, separating the real and the imaginary parts, we obtain 
	\begin{equation}\label{orthogonal}
		\left\{
		\begin{aligned}
			\sum_{i=1}^{n} (x_{ij}x_{ik}+y_{ij}y_{ik}) & =0, 1\leq j\neq k\leq n,  \\ 
			\sum_{i=1}^{n} (x_{ik}y_{ij}-x_{ij}y_{ik}) & =0, 1\leq j\neq k\leq n.
		\end{aligned}
		\right.
	\end{equation}
Note that the above equations are quadratic homogeneous polynomial equations. Then we can work over the polynomial ring $S=\mathbb{R}[x_{ij}, y_{ij}: 1\leq i,j\leq n]$. Therefore, any real solution of the above equation in $\mathbb R^{2n^2}$ corresponds to an orthogonal set of vectors in $\mathbb C^n$. In the language of algebraic varieties, we can say that any orthonormal set of $n$ vectors of $\mathbb C^n$ correspond to the real points of the variety $Z(I^{\{n\}})$, where $I^{\{n\}}$ is the ideal of the polynomial ring $S$ given by 

$I^{\{n\}}=(\{\sum_{i=1}^{n} (x_{ij}x_{ik}+y_{ij}y_{ik}), \sum_{i=1}^{n} (x_{ik}y_{ij}-x_{ij}y_{ik})|  1\leq j\neq k\leq n\})$.

Thus, the study of these ideals may be helpful while studying MUB, as any MUB has to satisfy the above equations and hence, must come from a point of the variety $Z(I^{\{n\}})$.
	
Suppose that, for a given dimension $n$, there is a set of $k$ MUBs, and we want to check whether this set of $k$ MUBs can be extended to a set of $k+1$ MUBs. This problem can also be interpreted in terms of the existence of solutions to a set of algebraic equations. Let us fix the following notations: Let $A_1, A_2, \ldots , A_k$ be a set of $k$ MUBs in dimension $n$. Let $A_l=\{\mathbf{a}_j^{(l)}: 1\leq j\leq n\}$, where $\mathbf{a}_j^{(l)}\in \mathbb C^n$, and $\mathbf{a}_j^{(l)}=(a_{1j}^{(l)}, a_{2j}^{(l)}, \ldots , a_{nj}^{(l)})^t, 1\leq j\leq n, 1\leq l\leq k$, where $a_{ij}^{(l)}\in \mathbb C$ are complex numbers. Now, we want to check whether there is a set of orthonormal vectors $A_{k+1}=\{\mathbf{a}_j^{(k+1)}: 1\leq j\leq n\}$, such that $A_1, A_2, \ldots , A_{k+1}$ is a system of $k+1$ MUBs. We now require to find out the algebraic relations that the vectors of the new MUB $A_{k+1}$ need to satisfy to be a part of the above system of MUBs. First, note that the vectors in $A_{k+1}$ have to be orthogonal, that is, they must come from the algebraic variety $Z(I^{\{n\}})$. More precisely, if we write $a^{(k+1)}_{ij}=b^{(k+1)}_{ij}+ic^{(k+1)}_{ij}$, then $x_{ij}= b^{(k+1)}_{ij}, y_{ij}=c^{(k+1)}_{ij}$ must be a solution of the system given in (\ref{orthogonal}). Moreover, they also need to satisfy the condition given in the definition of the MUB (Definition \ref{def_MUB}). In other words, the vectors $\mathbf{a}_j^{(k+1)}, 1\leq j \leq n$ must satisfy the following relations $|\langle\mathbf{a}_j^{(k+1)}\cdot \mathbf{a}_q^{(l)}\rangle|^2=\frac{1}{{n}}$ for all $1\leq j, q\leq n$, and $1\leq l\leq k$. This gives rise to the following relations:
	{\small
	\begin{align*}
		|\langle\mathbf{a}_j^{(k+1)}\cdot \mathbf{a}_q^{(l)}\rangle|^2-\frac{1}{{n}}
		&=|\sum_{i=1}^{n} a_{ij}^{(k+1)}\cdot \overline{a_{iq}^{(l)}}|^2-\frac{1}{{n}}\\ 
		&=|\sum_{i=1}^{n} (b_{ij}^{(k+1)}+ic_{ij}^{(k+1)})(b_{iq}^{(l)}-ic_{iq}^{(l)})|^2-\frac{1}{{n}}\\ 
		&= |\sum_{i=1}^{n} \{(b_{ij}^{(k+1)}b_{iq}^{(l)}+c_{ij}^{(l)}c_{iq}^{(l)}) + i(b_{iq}^{(l)}c_{ij}^{(k+1)}-b_{ij}^{(k+1)}c_{iq}^{(l)})\}|^2-\frac{1}{{n}}\\
		&=|\sum_{i=1}^{n} (b_{ij}^{(k+1)}b_{iq}^{(l)}+c_{ij}^{(k+1)}c_{iq}^{(l)}) + i\sum_{i=1}^{n} (b_{iq}^{(l)}c_{ij}^{(k+1)}-b_{ij}^{(k+1)}c_{iq}^{(l)})|^2-\frac{1}{{n}}\\
		&=\{\sum_{i=1}^{n} (b_{ij}^{(k+1)}b_{iq}^{(l)}+c_{ij}^{(k+1)}c_{iq}^{(l)})\}^2 + \{\sum_{i=1}^{n} (b_{iq}^{(l)}c_{ij}^{(k+1)}-b_{ij}^{(k+1)}c_{iq}^{(l)})\}^2-\frac{1}{{n}}.
	\end{align*}
	}
In other words, the vectors in the new MUB $A_{k+1}$ is a solution of the following system of algebraic equations in the variables $\mathbf{z_j}=(z_{1j}, z_{2j}, \ldots , z_{nj}), 1\leq j \leq n$, with $z_{ij}=x_{ij}+iy_{ij}$, where $x_{ij}, y_{ij}$ are real variables.
	\begin{equation}\label{MUB}
		\left\{
		\begin{aligned}
			&\sum_{i=1}^{n} (x_{ij}x_{ik}+y_{ij}y_{ik})  =0, 1\leq j\neq k\leq n,  \\ 
			&\sum_{i=1}^{n} (x_{ik}y_{ij}-x_{ij}y_{ik})  =0, 1\leq j\neq k\leq n, \\
			&\{\sum_{i=1}^{n} (x_{ij}b_{iq}^{(l)}+y_{ij}c_{iq}^{(l)})\}^2 + \{\sum_{i=1}^{n} (b_{iq}^{(l)}y_{ij}-x_{ij}c_{iq}^{(l)})\}^2 -\frac{1}{{n}}=0,\\ &1\leq j, q \leq n, 1\leq l\leq k.
		\end{aligned}
		\right.
	\end{equation}
Now, we consider the following ideals in the polynomial ring $S=\mathbb{R}[x_{ij}, y_{ij}: 1\leq i,j\leq n]$ for $1\leq l\leq k$: \[J_{l,k+1}^{\{n\}}=(\{\sum_{i=1}^{n} (x_{ij}b_{iq}^{(l)}+y_{ij}c_{iq}^{(l)})\}^2 + \{\sum_{i=1}^{n} (b_{iq}^{(l)}y_{ij}-x_{ij}c_{iq}^{(l)})\}^2 -\frac{1}{{n}}: 1\leq j,q\leq n).\]
Note that the ideal $J_{l,k+1}^{\{n\}}$ is generated by the polynomial relations coming from the MUB conditions between the pair of orthonormal bases $A_l$ and $A_{k+1}$ for all $1\leq l\leq k$. Now, we can give the following equivalent condition of whether a system of MUB of length $k$ in the vector space $\mathbb C^n$ can be extended to a system of MUB of length $k+1$.
\begin{theorem}
\label{real points}
Let $A_1, A_2, \ldots, A_k$ be a system of $k$ mutually unbiased  bases. Then this system can be extended to a system of $k+1$ MUBs if and only if the algebraic variety $Z(M_{k+1}^{\{n\}})$ of the ideal $M_{k+1}^{\{n\}}=I^{\{n\}}+ \sum_{i=1}^{k} J_{l,k+1}^{\{n\}}\subseteq \mathbb{R}[x_{ij}, y_{ij}: 1\leq i,j\leq n]$ has at least one point in $\mathbb R^{2n^2}$.
\end{theorem}
\begin{proof}
Suppose that the variety $Z(M_{k+1}^{\{n\}})$ has a point  $\alpha \in \mathbb R^{2n^2}$. Then, the point $\alpha$ corresponds to a real solution of the system (\ref{MUB}) given above. Hence, by identifying this solution using $z_{ij}=x_{ij}+iy_{ij}$, we get a new set of MUB $A_{k+1}=[z_{ij}]_{n\times n}$, which extends the given system of MUBs. Conversely, suppose that the given system can be extended to a system of $k+1$ MUBs. Then by the similar arguments given above, we can say that the algebraic variety $Z(M_{k+1}^{\{n\}})$ has at least one point in $\mathbb R^{2n^2}$. \qed
\end{proof}
As an immediate consequence, we have the following:
\begin{corollary}
Let $A_1, A_2,\ldots , A_k$ be a system of $k$ MUBs. If $M_{k+1}^{\{n\}}=(1)$, then this system of MUB cannot be extended to a system of $k+1$ MUBs.
\end{corollary}
Let us also present the following technical result.
\begin{proposition}
\label{first MUB}
Let $A_1, A_2,\ldots , A_k$ be a system of $k$ MUBs. Then for any unitary matrix $P$, the system $PA_1, PA_2, \ldots, PA_k$ is again a system of $k$ MUBs.
\end{proposition}
\begin{proof}
Note that for all $1\leq i\leq k$, the matrix $PA_i$ is again an unitary matrix, as \[(PA_i)(PA_i)^*= PA_iA_i^*P^*=PP^*=I_n.\] Since unitary matrices preserve inner products, it follows that the system $PA_1$, $PA_2$, $\ldots,$ $PA_k$ satisfies the inner product condition given in the definition of the MUBs. \qed
\end{proof}
Let us now consider the real MUBs, that is, only allowing real entries in Definition \ref{def_MUB}. In this case, the defining ideals of a system of MUB become much simpler, and can be obtained by putting $y_{ij}=0$ in the system of equations given in \ref{MUB}. Let $A_1, A_2, \ldots, A_k$ be a system of $k$ MUBs with real entries. Then the defining ideal of the condition that this system can be extended to a system of $k+1$ MUBs with real entries is given by \[M_{k+1}^{\{n\}}=I^{\{n\}}+ \sum_{i=1}^{k} J_{l,k+1}^{\{n\}}\subseteq \mathbb{R}[x_{ij}: 1\leq i,j\leq n],\]
where \[I^{\{n\}}=(\sum_{i=1}^{n} x_{ij}x_{ik} : 1\leq j\neq k\leq n),\] and \[J_{l,k+1}^{\{n\}}=(\sum_{i=1}^{n} x_{ij}b_{iq}^{(l)} -\frac{1}{{\sqrt{n}}}: 1\leq j,q\leq n).\]
We shall denote the defining ideals of a system of MUBs with real entries in the same way as of the complex case, and mention it explicitly whenever we use it.
	
\subsection{Ideals of the MUBs in dimension two}
\label{sec32}
In this subsection, we study ideals defining the MUBs in dimension 2 from the point of view of polynomial algebra. Although MUBs in dimension 2 are all known in terms of complex Hadamard matrices, we want to frame the problem of finding a new MUB in dimension 2 in terms of finding points in certain real algebraic varieties. In view of Proposition \ref{first MUB}, we may start with the identity matrix $I_2$, and first try to extend it to a MUB of length 2. We now continue our discussion in the following steps.
	
	\vspace{0.1cm}
	\noindent
	\textsc{\bf Step 1:}
	Let $E$ be the system of vectors coming from the identity matrix $I_2$. Then $A_{1}=\{\begin{pmatrix}1  \\ 0\end{pmatrix}, \begin{pmatrix}0 \\ 1\end{pmatrix}\}$, and we want to determine an orthonormal basis $A_2$ which makes the set $\{A_1,A_2\}$ a system of MUB of length 2.
\begin{proposition}
Let $E$ be the set of standard basis vectors of $\mathbb{C}^2$. Then any basis $B_1$ of $\mathbb{C}^2$, which makes the system $\{E,B_1\}$ a system of MUB, comes from a point on the real variety $Z(M_2^{\{2\}})$, where \[M_2^{\{2\}}=I^{\{2\}}+J_{1,4}^{\{2\}}, \]
		\[I^{\{2\}}=(x_{11}x_{12}+y_{11}y_{12}+ x_{21}x_{22}+y_{21}y_{22}, x_{11}y_{12}-x_{12}y_{11}+ x_{21}y_{22}-x_{22}y_{21}),\]
		\[J_{1,4}^{\{2\}}=(x_{11}^2+y_{11}^2-\frac{1}{2}, x_{12}^2+y_{12}^2-\frac{1}{2}, x_{21}^2+y_{21}^2-\frac{1}{2}, x_{22}^2+y_{22}^2-\frac{1}{2}).\]
	\end{proposition}
It is easy to check that $\{\frac{1}{\sqrt{2}}\begin{pmatrix}1 \\ 1\end{pmatrix}, \frac{1}{\sqrt{2}}\begin{pmatrix}1 \\-1\end{pmatrix}\}$ is a point on the variety $Z(M_2^{\{2\}})$ given above.
	
	\vspace{0.1cm}
	\noindent
	\textsc{\bf Step 2:}
	Let $B_1=\{\frac{1}{\sqrt{2}}\begin{pmatrix}1 \\ 1\end{pmatrix}, \frac{1}{\sqrt{2}}\begin{pmatrix}1 \\-1\end{pmatrix}\}$. then we have a system of MUB given by $\{E,B_1\}$, and we want to extend this to a system of MUB of length 3.
\begin{proposition}
Let $\{E,B_1\}$ be a system of MUB. Then any basis $B_2$ of $\mathbb{C}^2$, which makes the system $\{E,B_1, B_2\}$ a system of MUB, comes from a point on the real variety $Z(M_3^{\{2\}})$, where
		\[M_3^{\{2\}}=M_2^{\{2\}}+J_{2,4}^{\{2\}}, \]
		\begin{align*}
			J_{2,4}^{\{2\}}&=((x_{11}+x_{21})^2+(y_{11}+y_{21})^2-1,(x_{11}-x_{21})^2+(y_{11}-y_{21})^2-1,\\ & \hspace{1cm} (x_{12}+x_{22})^2+(y_{12}+y_{22})^2-1,(x_{12}-x_{22})^2+(y_{12}-y_{22})^2-1)
		\end{align*}
\end{proposition}
One can check that $\{\frac{1}{\sqrt{2}}\begin{pmatrix}1 \\ i\end{pmatrix}, \frac{1}{\sqrt{2}}\begin{pmatrix}1 \\-i\end{pmatrix}\}$ is a point on the variety $Z(M_3^{\{2\}})$ given above. Therefore, the collection $\{E,B_1,B_2\}$ is a system of MUBs for the vector space $\mathbb C^2$, where $B_{2}=\{\frac{1}{\sqrt{2}}\begin{pmatrix}1 \\ i\end{pmatrix}, \frac{1}{\sqrt{2}}\begin{pmatrix}1 \\-i\end{pmatrix}\}.$
	
It is known that one cannot extend the above system of MUB to a system of MUB of length 4. The discussions given below shows that our treatment of a system of MUB can detect this with simple algebraic manipulations.
\begin{proposition}
Let $\{E,B_1,B_2\}$ be a system of MUB in dimension 2 given by
		\begin{align*}
			E&=\{\begin{pmatrix}1  \\ 0\end{pmatrix}, \begin{pmatrix}0 \\ 1\end{pmatrix}\},\\
			B_{1}&=\{\frac{1}{\sqrt{2}}\begin{pmatrix}1 \\ 1\end{pmatrix}, \frac{1}{\sqrt{2}}\begin{pmatrix}1 \\-1\end{pmatrix}\},\\
			B_{2}&=\{\frac{1}{\sqrt{2}}\begin{pmatrix}1 \\ i\end{pmatrix}, \frac{1}{\sqrt{2}}\begin{pmatrix}1 \\-i\end{pmatrix}\}.
		\end{align*}
		Then this system cannot be extended to a system of MUB in dimension 2 of length greater than or equal to 4.
\end{proposition}
\begin{proof}
Let us assume there exists another MUB, say 
			$B_{3}=\{\begin{pmatrix}b_{12}+i c_{12}  \\ b_{21}+i c_{21}\end{pmatrix}, \begin{pmatrix} b_{12}+i c_{12} \\ b_{22}+ ic_{22}\end{pmatrix}\}$. Then it follows from Theorem \ref{real points} that $x_{ij}=b_{ij}, y_{ij}=c_{ij}, 1\leq i,j\leq 2$ is a point in the variety $Z(M_4^{\{2\}})$, where \[M_4^{\{2\}}= I^{\{2\}}+J_{1,4}^{\{2\}}+J_{2,4}^{\{2\}}+J_{3,4}^{\{2\}}.\]
			Now we claim that $M_4^{\{2\}}=(1)$ in $R=\mathbb K[x_{ij}, y_{ij}, 1\leq i,j\leq 2]$. Note that 
			
			\begin{align*}
				I^{\{2\}}&=(x_{11}x_{12}+y_{11}y_{12}+ x_{21}x_{22}+y_{21}y_{22}, x_{11}y_{12}-x_{12}y_{11}+ x_{21}y_{22}-x_{22}y_{21})\\
				J_{1,4}^{\{2\}}&=(x_{11}^2+y_{11}^2-\frac{1}{2}, x_{12}^2+y_{12}^2-\frac{1}{2}, x_{21}^2+y_{21}^2-\frac{1}{2}, x_{22}^2+y_{22}^2-\frac{1}{2})\\
				J_{2,4}^{\{2\}}&=((x_{11}+x_{21})^2+(y_{11}+y_{21})^2-1,(x_{11}-x_{21})^2+(y_{11}-y_{21})^2-1,\\ & \hspace{1cm} (x_{12}+x_{22})^2+(y_{12}+y_{22})^2-1,(x_{12}-x_{22})^2+(y_{12}-y_{22})^2-1)\\
				J_{3,4}^{\{2\}}&=((x_{11}+y_{21})^2+(y_{11}-x_{21})^2-1, (x_{11}-y_{21})^2+(y_{11}+x_{21})^2-1, \\ & \hspace{1cm} (x_{12}+y_{22})^2+(y_{12}-x_{22})^2-1, (x_{12}-y_{22})^2+(y_{12}+x_{22})^2-1)
			\end{align*}
			Now, we do the following simplifications in the quotient ring $R/ M_4^{\{2\}}$. For our convenience, we shall write $f$ instead of $\overline{f}$ to denote the residue class of an element $f\in R$ in the quotient ring $R/ M_4^{\{2\}}$.
			\begin{align*}
				(x_{11}+x_{21})^2+(y_{11}+y_{21})^2-1 &= x_{11}^2+x_{21}^2+y_{11}^2+y_{21}^2+ 2(x_{11}x_{21}+y_{11}y_{21})-1\\ & = \frac{1}{2} + \frac{1}{2}+ 2(x_{11}x_{21}+y_{11}y_{21})-1\\  &=2(x_{11}x_{21}+y_{11}y_{21})
			\end{align*}
			Similarly, we get 
			\begin{align*}
				(x_{11}+y_{21})^2+(y_{11}-x_{21})^2-1= 2(x_{11}y_{21}- y_{11}x_{21})
			\end{align*}
			
			Using the above two relations, and the relations given in the ideal $J_{1,4}^{\{2\}}$, we obtain
			{\small 
			\begin{align*}
				& (x_{11}x_{12}+y_{11}y_{12}+ x_{21}x_{22}+y_{21}y_{22})^2+ (x_{11}y_{12}-x_{12}y_{11}+ x_{21}y_{22}-x_{22}y_{21})^2\\
                &=  x_{11}^2x_{12}^2+y_{11}^2y_{12}^2+ x_{21}^2x_{22}^2+y_{21}^2y_{22}^2+ 2(x_{11}x_{12}y_{11}y_{12}+ x_{11}x_{12} x_{21}x_{22}  \\ & \hspace{0.5cm} +x_{11}x_{12}y_{21}y_{22}+ y_{11}y_{12}x_{21}x_{22} + y_{11}y_{12}y_{21}y_{22}+ x_{21}x_{22}y_{21}y_{22})+ x_{11}^2y_{12}^2+x_{12}^2y_{11}^2 \\ & \hspace{0.5cm} +x_{21}^2y_{22}^2+x_{22}^2y_{21}^2 +2(-x_{11}y_{12}x_{12}y_{11} +x_{11}y_{12}x_{21}y_{22}- x_{11}y_{12}x_{22}y_{21} \\ & \hspace{0.5cm} -x_{12}y_{11} x_{21}y_{22}+ x_{12}y_{11}x_{22}y_{21}- x_{21}y_{22}x_{22}y_{21})\\&=
				x_{11}^2(x_{12}^2+ y_{12}^2)+ y_{11}^2(y_{12}^2+ x_{12}^2)+x_{21}^2(x_{22}^2+y_{22}^2)+ y_{21}^2(y_{22}^2+x_{22}^2) + \\ & \hspace{0.5cm} 2(x_{11}x_{12} x_{21}x_{22} + x_{11}x_{12}y_{21}y_{22}+ y_{11}y_{12}x_{21}x_{22} + y_{11}y_{12}y_{21}y_{22} +x_{11}y_{12}x_{21}y_{22}\\ & \hspace{0.5cm} - x_{11}y_{12}x_{22}y_{21}  -x_{12}y_{11} x_{21}y_{22}+ x_{12}y_{11}x_{22}y_{21})\\ &= 
				\frac{1}{2}x_{11}^2+ \frac{1}{2}y_{11}^2+ \frac{1}{2}x_{21}^2+ \frac{1}{2}y_{21}^2 + 2\{x_{11}x_{21} (x_{12}x_{22}+ y_{12}y_{22}) + x_{11}y_{21}(x_{12}y_{22}- y_{12}x_{22}) \\ & \hspace{0.5cm} + y_{11}x_{21}(y_{12}x_{22} -x_{12}y_{22} ) + y_{11}y_{21}(y_{12}y_{22}+ x_{12}x_{22}) \}\\&= 
				\frac{1}{2}\cdot \frac{1}{2}+ \frac{1}{2}\cdot \frac{1}{2} + 2\{(x_{11}x_{21}+ y_{11}y_{21}) (x_{12}x_{22}+ y_{12}y_{22})\\  & \hspace{5cm}+ (x_{11}y_{21}-y_{11}x_{21} )(x_{12}y_{22}- y_{12}x_{22})  \} = \frac{1}{2}.
			\end{align*}
			}
This shows that $\frac{1}{2}\in M_4^{\{2\}}$, and hence $M_4^{\{2\}}=(1)$ in $R$. Therefore, the given system of MUB cannot be extended to a system of MUB of length $4$. \qed
		\end{proof}

\subsection{Results On Complete Intersection}
\label{sec33}
In this subsection, we study the ideals arising from the MUBs from a view of commutative algebra. In the following theorem, we show that the ideal $J_{1,n}^{\{m\}}$ is a prime ideal generated by a regular sequence.
\begin{theorem}
\label{sphere}
Let  $A_1,A_2, \ldots , A_{m-1}$ be a system of MUBs, and assume that $A_1=I_n$, the identity matrix. Then  $J_{1,m}^{\{n\}}$ is a complete intersection prime ideal.
\end{theorem}
\begin{proof}
Let $J_{1,m}^{\{n\}}=(f_1,\ldots , f_{n^2})$. Note that $lcm(in(f_i),in(f_j))=in(f_i)in(f_j)$ for all $i\neq j$, and hence by Lemma \ref{lcm} the S-polynomials $S(f_i,f_j)$ reduces to $0$ with respect to $f_1,\ldots , f_{m^2}$. Therefore, by Theorem \ref{BBC}, $\{f_1, \ldots , f_{n^2}\}$ is a Gr\"obner basis of $J_{1,m}^{\{n\}}$. Note that $(in(f_1), \ldots , in(f_{n^2}))$ is complete intersection ideal, as $\{in(f_i), 1\leq i\leq n^2\}$ are monomials with $Supp(f_i)\cap Supp(f_j)=\phi$ for all $i\neq j$. Therefore, $J_{1,m}^{\{n\}}$ is also a complete intersection ideal.

To show that $R/J_{1,m}^{\{n\}}$ is a domain, we consider the following embedding of rings: \[S/J_{1,m}^{\{n\}}= \frac{\mathbb{R}[x_{ij}, y_{ij}: 1\leq i,j\leq n]}{J_{1,m}^{\{n\}}} \hookrightarrow \frac{\mathbb{C}[x_{ij}, y_{ij}: 1\leq i,j\leq n]}{J_{1,m}^{\{n\}}} ,\] where the embedding is defined by sending any $r\in \mathbb R$ to $r\in \mathbb{C}$, and $\overline{x_{ij}}, \overline{y_{ij}}\in S/J_{1,m}^{\{n\}}$ to $\overline{x_{ij}}, \overline{y_{ij}}\in \frac{\mathbb{C}[x_{ij}, y_{ij}: 1\leq i,j\leq n]}{J_{1,m}^{\{n\}}}$ respectively. If we can show that the ring $\frac{\mathbb{C}[x_{ij}, y_{ij}: 1\leq i,j\leq n]}{J_{1,m}^{\{n\}}}$ is a domain, then we are through. As $J_{1,m}^{\{n\}}=(x_{ij}^2+ y_{ij}^2-\frac{1}{n}: 1\leq i,j\leq n)$, we can write  
\[\frac{\mathbb{C}[x_{ij}, y_{ij}: 1\leq i,j\leq n]}{J_{1,m}^{\{n\}}}=\bigotimes_{1\leq i,j\leq n} \frac{\mathbb{C}[x_{ij}, y_{ij}]}{(x_{ij}^2+ y_{ij}^2-\frac{1}{n})}.\]
Now, we shall show that $\frac{\mathbb{C}[x_{ij}, y_{ij}]}{(x_{ij}^2+ y_{ij}^2-\frac{1}{n})}$ is an integral domain for all $1\leq i,j\leq n$. Note that there is an isomorphism \[\frac{\mathbb{C}[U_{ij}, V_{ij}]}{(U_{ij}V_{ij}-1)}\cong \frac{\mathbb{C}[x_{ij}, y_{ij}]}{(x_{ij}^2+ y_{ij}^2-\frac{1}{n})}\] by identifying $U_{ij}\mapsto x_{ij}+i y_{ij}$ and  $V_{ij}\mapsto x_{ij}-i y_{ij}$. But the ring $\frac{\mathbb{C}[U_{ij}, V_{ij}]}{(U_{ij}V_{ij}-1)}$ is a localization of the polynomial ring $\mathbb{C}[U_{ij}]$ at the multiplicative-ly closed set $\{1,U_{ij}, U_{ij}^2,\ldots \}$. Therefore the ring $\frac{\mathbb{C}[U_{ij}, V_{ij}]}{(U_{ij}V_{ij}-1)}$, being localization of an integral domain, is also an integral domain. As tensor product of integral domains is again an integral domain, we conclude that $\bigotimes_{1\leq i,j\leq n} \frac{\mathbb{C}[x_{ij}, y_{ij}]}{(x_{ij}^2+ y_{ij}^2-\frac{1}{n})}$ is an integral domain. Therefore, the ring $\frac{\mathbb{C}[x_{ij}, y_{ij}: 1\leq i,j\leq n]}{J_{1,m}^{\{n\}}}$ is an integral domain, and hence the ideal $J_{1,m}^{\{n\}}$ is a prime ideal generated by a regular sequence. \qed
\end{proof}

As an immediate consequence, we have the following
\begin{corollary}
For any $m, n\in \mathbb{N}$, $ht(J_{1,m}^{\{n\}})=n^2$.
\end{corollary}
\begin{proof}
Since $J_{1,m}^{\{n\}}$ is generated by a regular sequence in a polynomial ring, by \cite[Theorem 17.4.]{Matsumura1987} we have $ht(J_{1,m}^{\{n\}})=n^2$. \qed
\end{proof}

The next result shows that the ideal $J_{1,m}^{\{n\}}$ coming from the system of MUBs with real entries is a prime ideal generated by a regular sequence.
\begin{theorem}
Let  $A_1, A_2, \ldots, A_{m-1}$ be a system of MUB, and assume that $A_1=I_n$, the identity matrix. Then $J_{1,m}^{\{n\}}$ is a maximal ideal generated by a regular sequence, where $J_{1,m}^{\{n\}}$ is the ideal coming from the system of MUBs with real entries, given by \[J_{1,m}^{\{n\}}=(x_{ij} -\frac{1}{{\sqrt{n}}}: 1\leq i,j\leq n)\]
\end{theorem}
\begin{proof}
Observe that $\frac{\mathbb{R}[x_{ij}: 1\leq i,j\leq n]}{(x_{ij} -\frac{1}{{\sqrt{n}}}: 1\leq i,j\leq n)}\cong \mathbb R$, and hence the assertion holds. \qed
\end{proof}
We conclude this section with an open-ended question as conjecture that came out of our study of the defining ideals of a system of MUBs.
\begin{conjecture}
Let  $A_1, A_2, \ldots, A_{m-1}$ be a system of MUB, and assume that $A_1=I_n$, the identity matrix. Then the ideal  $I^{\{n\}}+J_{1,m}^{\{n\}}$ is a prime ideal generated by a regular sequence. Moreover, $ht(I^{\{n\}}+J_{1,m}^{\{n\}})=2n^2-n$.
\end{conjecture}

\section{Maximal Commuting Bases and MUBs }
\label{sec4}
The question that we are discussing in this initiative is related to examine algebraic results related to MUBs. In this regard, we consider extending the result of \cite[Theorems 3.2 and 3.4]{Bandyopadhyay} under the hypothesis related to existence of normal matrices. We prove that  there is an one-to-one correspondence between the MUBs and the maximal commuting classes (bases) of orthogonal normal matrices in $C^{n}$. It means that for $m$ MUBs in $C^n$, there are $m$ commuting classes each consisting of $n$ commuting orthogonal normal matrices. The existence of maximal commuting basis for $\mathcal M_n({\mathbb{C}})$ ensures the complete set of MUBs in $\mathcal M_n({\mathbb{C}})$.

Let $\mathcal M_n({\mathbb{C}})$ denote the set of all $n\times n$ complex matrices. For any matrix $A\in \mathcal M_n({\mathbb{C}})$, we use the standard notation $A^{\dagger}$ to denote the transpose of the conjugate matrix $A$. Two matrices $A,B\in\mathcal M_n({\mathbb{C}})$ are orthogonal iff their trace inner product $<A,B>=tr(A^{\dagger}B)=0.$

An $n\times n$ complex matrix $A \in \mathcal M_n({\mathbb{C}})$ is called \emph{unitarily diagonalizable} if $U^{\dagger}AU$ is diagonal for some unitary matrix $U$. Let us now present a few technical results.

\begin{lemma}
\label{unitary diagonalizable}
An $n\times n$ complex matrix $A$ is unitarily diagonalizable if and only if $A$ is normal.
\end{lemma}

\begin{lemma}\label{max orthogonal}
There are at most $n$ pairwise orthogonal commuting normal matrices in $\mathcal M_n({\mathbb{C}})$.
\end{lemma}
\begin{proof}
Let $A_1, \ldots, A_m$ be pairwise orthogonal commuting normal matrices in $\mathcal M_n({\mathbb{C}})$. From lemma ~\ref{unitary diagonalizable}, there exists a unitary matrix, say $U\in \mathcal M_n({\mathbb{C}})$, such that they are simultaneously diagonalizable. Then, the matrices  $U^*A_1U, \ldots, U^*A_mU$ is a collection diagonal orthogonal matrices. In this way, we get $m$ orthogonal vectors which are basically the diagonal of $U^{\dagger}A_iU$ for $i=1, 2, \ldots, m$. Therefore, $m \leq n$. \qed
\end{proof}
Let $\mathcal{B} = \{U_1, \ldots, U_{n^2}\}$ be a basis of normal matrices of $\mathcal M_n({\mathbb{C}})$ such that $U_1$ is the identity matrix $I_n\in \mathcal M_n({\mathbb{C}})$. We say that the basis $\mathcal B$ is a \textit{maximal commuting basis} for $\mathcal M_n({\mathbb{C}})$ if B can be partitioned as
\[
\mathcal{B}=\{I_n\}\cup \mathcal{C}_1\cup \cdots \mathcal{C}_{n+1}
\] 
where each contains exactly $n-1$ commuting matrix from $\mathcal{B}$. In the next two theorems, we extend the result of \cite[Theorems 3.2 and 3.4]{Bandyopadhyay} under the hypothesis of existence of normal matrices.
\begin{theorem}
If there is a maximal commuting basis of orthogonal normal matrices in $\mathcal M_n({\mathbb{C}})$, then there is a set of $n+1$ mutually unbiased bases.
\end{theorem}
\begin{proof}
Let $\mathcal{B}$ be a maximal commuting basis of orthogonal unitary matrices in $\mathcal M_n({\mathbb{C}})$. Consider the decomposition of $\mathcal{B}$ as follows:
    \[
    \mathcal{B}=\{I_n\}\cup \mathcal{C}_1\cup \cdots \cup \mathcal{C}_{n+1}.
    \]
    For $1\leq i\leq n+1$, we set 
    \[
    \mathcal{C}'_i=\{I_n\}\cup \mathcal{C}_i=\{U_{i,0},U_{i,1},\ldots ,U_{i,n-1}\}.
    \]
    Then by Lemma~\ref{max orthogonal}, each $\mathcal{C}'_i$ is a maximal set of pairwise orthogonal commuting normal matrices in $\mathcal M_n({\mathbb{C}})$. Then for each $1\leq i\leq n+1$, there is a unitary matrix $V_i$ such that $V_i^*U_{i,t}V_i$ are diagonal matrices for each $0\leq t\leq n-1$. Let $V_i=\{\ket{v_{1}^i},\ket{v_{2}^i},...,\ket{v_{n}^i}\}$, where $\ket{v_{k}^i}$ are the column vectors of $V_i$. Then for all $0\leq t\leq n-1$, 
    \[
    U_{i,t}=\sum_{k=1}^{n} \lambda_{i,t,k} \ket{v_{k}^i} \bra{v_{k}^i}
    \]
We claim that, the matrices $V_1,\ldots , V_{n+1}$ form a system of MUBs. It suffices to show that $|\braket{v_{k}^i \mid v_{l}^j}|^2=\frac{1}{n}$ for all $1\leq i,j\leq n+1$ and $1\leq k,l\leq n$. Now, due to the orthogonality of the collection $\mathcal{B}$, it follows that 
    \begin{align*}
        \mathrm{Tr}(U_{i,p}^*U_{j,q})=
        \begin{cases}
            n &\text{ if $p=q=0$},\\
            0 &\text{ otherwise}.
        \end{cases}
    \end{align*}
    On the other hand, 
    \begin{align*}
        \mathrm{Tr}(U_{i,p}^*U_{j,q})&=\mathrm{Tr}((\sum_{k=1}^{n} \lambda_{i,p,k} \ket{v_{k}^i} \bra{v_{k}^i})^* \sum_{l=1}^{n} \lambda_{j,q,l} \ket{v_{l}^j} \bra{v_{l}^j})\\
        &=\sum_{k=1}^{n}\sum_{l=1}^{n}  \lambda_{i,p,k}^*\lambda_{j,q,l} \mathrm{Tr}((\ket{v_{k}^i} \bra{v_{k}^i})^* (\ket{v_{l}^j} \bra{v_{l}^j}))\\
        &=\sum_{k=1}^{n}\sum_{l=1}^{n}  \lambda_{i,p,k}^*\lambda_{j,q,l} |\braket{v_{k}^i \mid v_{l}^j}|^2.
    \end{align*}
    Therefore, comparing the above, we obtain for all $1\leq i,j\leq n+1$ and $0\leq p,q\leq n-1$,
    \[
    \sum_{k=1}^{n}\sum_{l=1}^{n}  \lambda_{i,p,k}^*\lambda_{j,q,l} |\braket{v_{k}^i \mid v_{l}^j}|^2=\begin{cases}
            n &\text{ if $p=q=0$},\\
            o &\text{ otherwise}.
        \end{cases}
    \]
    For each $1\leq i\leq n+1$, consider the matrices $M_i$ whose $k^{\text{th}}$ row is the diagonal entries of the matrix $V_iU_{i,k}V_i^*$. Then $M_i$ is a unitary matrix for all $1\leq i\leq n+1$. Note that the above equality can ve written as $AP=\lambda$, where
    \begin{align*}
         A&=M_i^*\otimes M_j\\
         P&=(\braket{v_{1}^i \mid v_{1}^j}|^2, \braket{v_{2}^i \mid v_{2}^j}|^2,\ldots , \braket{v_{n}^i \mid v_{n}^j}|^2)^{T}\\
         \lambda &=(n,0,0,\ldots ,0)^T
    \end{align*}
    Since $A$ is again an unitary matrix whose first row is the vector $(1, 1, \ldots, 1)$, it follows that  
    $\braket{v_{k}^i \mid v_{l}^j}|^2=\frac{1}{n}$ for $1\leq k, l \leq n$. \qed    
\end{proof}
Now we require a technical result.
\begin{lemma}
Consider $\mathbb{C}^n$, let $\{1 = v_1, v_2, \ldots, v_n\}$ be the orthonormal basis vector of $\mathbb{C}^n$ where $1=\frac{1}{\sqrt{n}}\begin{bmatrix}
         1 &   1 & 1 & \ldots  &1
        \end{bmatrix}^{t}=v_1$ . If $v_j=\{v_{j1},v_{j2},...,v_{jn}\}$ then $<v_1,v_j>=0,$ that means $\sum_iv_{ji}=0$;  $\forall j = 2, 3, \ldots, n.$
\end{lemma}
Using the above lemma, we prove the following result.
\begin{theorem}
\label{MUB construction}
    Let $B_1, B_2, \ldots, B_m$ be a set of MUBs in  $\mathbb{C}^n$. Then there are $m$ classes $C_1, C_2, \ldots, C_m$ each consisting of $n$ commuting normal matrices, such that matrices in $C_1 \cup C_2 \cup \ldots \cup C_m$ are pairwise orthogonal.
\end{theorem}
\begin{proof}
    Let $B_j=\{\ket{\psi^j_1}, \ket{\psi^j_2}, \ldots, \ket{\psi^j_n}\}$. Then 
    \begin{align*}
       \braket{\psi^j_s|\psi^j_t}=\delta_{s,t}; 1\leq s,t\leq n. \\|\braket{\psi^j_s|\psi^j_t}|^2=\frac{1}{n}, 1\leq j <k\leq m; 1\leq s,t\leq n. 
    \end{align*}
   We label the matrices in the class $C_j$ as $C_j=\{U_{j1}, U_{j2}, \ldots,U_{jn}\}$ where
\[ U{jt}=\sum^n_{i=1}v_{ti}\ket{\psi^j_i}\bra{\psi^j_i}; t=1,2, \ldots, n.\] 

Note that $U_{j1}=I_n$ for $j=1, 2, \ldots, m$ and $U_{js}, U_{jt}$ are commuting, because both are diagonal relative to basis $B_j$. Finally $U_{js}$ are normal matrices as 
\[
U_{js} U^{\dagger}_{js} = \left( \sum_{k=1}^n v_{sk} \ket{\psi^j_k} \bra{\psi^j_k} \right)
\left( \sum_{i=1}^n v_{si} \ket{\psi^j_i} \bra{\psi^j_i} \right)^\dagger
\]

\[
= \sum_{k,i=1}^n v_{sk} \, \overline{v_{si}} \, \ket{\psi^j_k} \braket{\psi^j_k | \psi^j_i} \bra{\psi^j_i}
= \sum_{k=1}^n |v_{sk}|^2 \ket{\psi^j_k} \bra{\psi^j_k}
= U^{\dagger}_{js} U_{js}
\]

Now we show that any pair of matrices $U_{js},U_{kj}$ are orthogonal if $s=j\neq 1$ in [consider $(js)\neq (kt)$].
\begin{align*}
\langle U_{js}, U_{kt} \rangle &= \operatorname{Tr}(U^{\dagger}_{js} U_{kt}) \\
&= \operatorname{Tr} \left( \left( \sum_{m=1}^n v_{sm} \ket{\psi^j_m} \bra{\psi^j_m} \right)^{\dagger}
\left( \sum_{i=1}^n v_{ti} \ket{\psi^k_i} \bra{\psi^k_i} \right) \right) \\
&= \operatorname{Tr} \left( \sum_{m=1}^n \sum_{i=1}^n v_{ti} v^{\dagger}_{sm}
\ket{\psi^j_m} \bra{\psi^j_m} \braket{\psi^j_m | \psi^k_i} \bra{\psi^k_i} \right)
\end{align*}
Now,
\[
\braket{\psi^j_m | \psi^k_i} = 
\begin{cases}
\delta_{mi}, & \text{if } j = k, \\
\frac{1}{\sqrt{n}}, & \text{if } j \neq k
\end{cases}
\]
Thus,
\begin{align*}
\langle U_{js}, U_{kt} \rangle &= \operatorname{Tr} \left( \sum_{m,i} v^{\dagger}_{sm} v_{ti} \ket{\psi^j_m} 
\braket{\psi^j_m | \psi^k_i} \bra{\psi^k_i} \right) \\
&= \sum_{m,i} v^{\dagger}_{sm} v_{ti} \operatorname{Tr} \left( 
\ket{\psi^j_m} \braket{\psi^j_m | \psi^k_i} \bra{\psi^k_i} \right) \\
&= \sum_{m,i} v^{\dagger}_{sm} v_{ti} \left| \braket{\psi^j_m | \psi^k_i} \right|^2
\end{align*}
If \( j = k \), then
$
\langle v_{js}, v_{kt} \rangle = \sum_m v^{\dagger}_{sm} v_{tm} = \langle v_s, v_t \rangle = 0
$.

\noindent If \( j \neq k \), then
$\langle v_{js}, v_{kt} \rangle = \sum_{m,i} v^{\dagger}_{sm} v_{ti} \cdot \frac{1}{n}
= \frac{1}{n} \sum_{m,i} v^{\dagger}_{sm} v_{ti}
= \frac{1}{n} \left( \sum_m v^{\dagger}_{sm} \right) \left( \sum_i v_{ti} \right)
= \frac{1}{n} \times 0 \times 0 = 0$.
\qed
\end{proof}

We now present an example for $n = 4$. There are five mutually unbiased bases (MUBs), denoted by $(I, B_2, B_3, B_4, B_5)$, in dimension $n = 4$. Hence, we have five commuting classes $(C_1, C_2, C_3, C_4, C_5)$, each containing four orthogonal normal matrices. Let $B = \{1 = v_1, v_2, v_3, v_4\}$ be an orthonormal basis of $\mathbb{C}^4$, where each basis vector $v_j = \{v_{j1}, v_{j2}, v_{j3}, v_{j4}\}$.
Here,
\[
\begin{aligned}
1=v_1 &= \frac{1}{2}(1, 1, 1, 1)^\dagger, \\
v_2 &= \frac{1}{2}(1, 0, -1, 0)^\dagger, \\
v_3 &= \frac{1}{2}(0, 1, 0, -1)^\dagger, \\
v_4 &= \frac{1}{2}(1, -1, 1, -1)^\dagger.
\end{aligned}
\]
These are five MUBs for $d = 4$ as follows.
\[
\begin{aligned}
B_1=I_4 &= \begin{bmatrix}
1 & 0 & 0 & 0 \\
0 & 1 & 0 & 0 \\
0 & 0 & 1 & 0 \\
0 & 0 & 0 & 1
\end{bmatrix}, \quad
B_2 = \frac{1}{2} \begin{bmatrix}
1 & 1 & 1 & 1 \\
1 & 1 & -1 & -1 \\
1 & -1 & -1 & 1 \\
1 & -1 & 1 & -1
\end{bmatrix}, \quad
B_3 = \frac{1}{2} \begin{bmatrix}
1 & 1 & 1 & 1 \\
-1 & -1 & 1 & 1 \\
-i & i & i & -i \\
-i & i & -i & i
\end{bmatrix}, \\
B_4 &= \frac{1}{2} \begin{bmatrix}
1 & 1 & 1 & 1 \\
-i & -i & i & i \\
-i & i & i & -i \\
-1 & 1 & -1 & 1
\end{bmatrix}, \quad
B_5 = \frac{1}{2} \begin{bmatrix}
1 & 1 & 1 & 1 \\
-i & -i & i & i \\
-1 & 1 & -1 & 1 \\
-i & i & i & -i
\end{bmatrix}.
\end{aligned}
\]
Following Theorem~\ref{MUB construction}, we obtain the following five (maximal) commuting basis $(C_1, C_2, C_3, C_4, C_5)$, each containing four orthogonal normal commuting matrices and the first one is identity.
\[
\begin{aligned}
C_1 &= \left\{
\begin{bmatrix}
1 & 0 & 0 & 0 \\
0 & 1 & 0 & 0 \\
0 & 0 & 1 & 0 \\
0 & 0 & 0 & 1
\end{bmatrix},
\quad
\begin{bmatrix}
1 & 0 & 0 & 0 \\
0 & 0 & 0 & 0 \\
0 & 0 & -1 & 0 \\
0 & 0 & 0 & 0
\end{bmatrix},
\quad
\begin{bmatrix}
0 & 0 & 0 & 0 \\
0 & 1 & 0 & 0 \\
 0 & 0 & 0 & 0\\
0 & 0 & 0 & -1
\end{bmatrix},
\quad
\begin{bmatrix}
1 & 0 & 0 & 0 \\
0 & -1 & 0 & 0 \\
 0 & 0 & 1 & 0\\
0 & 0 & 0 & -1
\end{bmatrix}
\right\}, \\[1.5ex]
C_2 &= \left\{
\begin{bmatrix}
1 & 0 & 0 & 0 \\
0 & 1 & 0 & 0 \\
0 & 0 & 1 & 0 \\
0 & 0 & 0 & 1
\end{bmatrix}, \quad \frac{1}{4}
\begin{bmatrix}
0 & 2 & 2 & 0 \\
2 & 0 & 0 & 2 \\
2 & 0 & 0 & 2 \\
0 & 2 & 2 & 0
\end{bmatrix}, \quad \frac{1}{4}
\begin{bmatrix}
    0 & 2 & -2 & 0 \\
    2 & 0 & 0 & -2 \\
    -2 & 0 & 0 & 2 \\
    0 & -2 & 2 & 0
\end{bmatrix}, \quad 
\begin{bmatrix}
    0 & 0 & 0 & 1 \\
    0 & 0 & 1 & 0 \\
    0 & 1 & 0 & 0 \\
    1 & 0 & 0 & 0
\end{bmatrix},
\right\}, \\[1.5ex]
C_3 &= \left\{
\begin{bmatrix}
1 & 0 & 0 & 0 \\
0 & 1 & 0 & 0 \\
0 & 0 & 1 & 0 \\
0 & 0 & 0 & 1
\end{bmatrix}, \quad \frac{1}{4}
\begin{bmatrix}
    0 & 2 & 2i & 0 \\
    -2 & 0 & 0 & -2i \\
    -2i & 0 & 0 & 2 \\
    0 & 2i & 2 & 0
\end{bmatrix}, \quad \frac{1}{4}
\begin{bmatrix}
    0 & -2 & -2i & 0\\
    -2 & 0 & 0 & 2i \\
    2i & 0 & 0 & 2 \\
    0 & -2i & 2 & 0
\end{bmatrix}, \quad 
\begin{bmatrix}
    0 & 0 & 0 & i \\
    0 & 0 & -i & 0 \\
    0 & i & 0 & 0 \\
    -i & 0 & 0 & 0
\end{bmatrix}
\right\}, \\[1.5ex]
C_4 &= \left\{
\begin{bmatrix}
1 & 0 & 0 & 0 \\
0 & 1 & 0 & 0 \\
0 & 0 & 1 & 0 \\
0 & 0 & 0 & 1
\end{bmatrix}, \quad \frac{1}{4}
\begin{bmatrix}
    0 & 2i & 2i & 0 \\
    -2i & 0 & 0 & 2i \\
    -2i & 0 & 0 & 2i \\
    0 & -2i & -2i & 0 \\
\end{bmatrix}, \quad \frac{1}{4}
\begin{bmatrix}
    0 & 2i & -2i & 0\\
    -2i & 0 & 0 & -2i \\
    2i & 0 & 0 & 2i \\
    0 & 2i & -2i & 0
\end{bmatrix}, \quad 
\begin{bmatrix}
    0 & 0 & 0 & -1 \\
    0 & 0 & 1 & 0 \\
    0 & 1 & 0 & 0 \\
    -1 & 0 & 0 & 0
\end{bmatrix}
\right\}, \\[1.5ex]
C_5 &= \left\{
\begin{bmatrix} 
1 & 0 & 0 & 0 \\
0 & 1 & 0 & 0 \\
0 & 0 & 1 & 0 \\
0 & 0 & 0 & 1
\end{bmatrix}, \quad \frac{1}{4}
\begin{bmatrix}
    0 & 2i & 0 & 2i \\
    -2i & 0 & 2i & 0 \\
    0 & -2i & 0 & -2i \\
    -2i & 0 & 2i & 0
\end{bmatrix}, \quad \frac{1}{4}
\begin{bmatrix}
    0 & 2i & 0 & -2i\\
    -2i & 0 & -2i & 0 \\
    0 & 2i & 0 & -2i \\
    2i & 0 & 2i & 0
\end{bmatrix}, \quad 
\begin{bmatrix}
    0 & 0 & -1 & 0 \\
    0 & 0 & 0 & 1 \\
    -1 & 0 & 0 & 0 \\
    0 & 1 & 0 & 0
\end{bmatrix}
\right\}.
\end{aligned}
\]
This concludes the example.

\section{Conclusion}
As it is well known, the problem of extending a system of MUBs is in general a difficult problem. Our approach to solving this essentially boils down to finding real solutions to some specific system of polynomial equations over $\mathbb R$. Our work presents relevant theoretical study in this direction for further insight. In general, the main difficulty is the complexity of the Gr\"obner basis algorithm, which is very high in general. However, our result regarding complete intersection demonstrated that it is possible to reduce the number of equations using techniques from commutative algebra. We have also shown that there is a one-to-one correspondence between MUBs and the maximal commuting classes (bases) of orthogonal normal matrices.


\begin{thebibliography}{100}

\bibitem{Bandyopadhyay} 
S. Bandyopadhyay, P.O. Boykin, V. Roychowdhury and F. Vatan. 
\textit{A New Proof for the Existence of Mutually Unbiased Bases.} 
Algorithmica 34, 512–528 (2002). https://doi.org/10.1007/s00453-002-0980-7

\bibitem{baumgratz2014quantifying} 
T. Baumgratz, M. Cramer and M. B. Plenio. 
\textit{Quantifying coherence.} 
Physical review letters, 113(14), 140401, (2014).

\bibitem{BB84}
C. H. Bennett and G. Brassard. 
\textit{Quantum cryptography: Public key distribution and coin tossing.} 
Theoretical computer science, 560, 7-11 (2014).

\bibitem{bengtsson} 
I. Bengtsson and A. Ericsson. 
\textit{MUBs and the complementarity polytope.} 
Open Sys. Inf. Dyn. 12, 107 (2005).
doi: \url{https://doi.org/10.1007/s11080-005-5721-3}

\bibitem{boykin} 
P. O. Boykin, M. Sitharam, P. H. Tiep, and P. Wocjan. 
\textit{MUBs and orthogonal decompositions of Lie algebras.} Quant. Inform. Comput. 7, 371 (2007).
doi: \url{https://doi.org/10.26421/QIC7.4-6}

\bibitem{brierley2008maximal} 
S. Brierley and S. Weigert. 
\textit{Maximal sets of mutually unbiased quantum states in dimension 6}.
Physical Review A, (2008).
doi: \url{10.1103/physreva.78.042312}

\bibitem{brierley2010mutually} 
S. Brierley and S. Weigert. 
\textit{Mutually unbiased bases and semi-definite programming}.
Journal Of Physics: Conference Series. \textbf{254}, 012008 (2010)

\bibitem{Weigert} 
S. Brierley, S. Weigert and I. Bengtsson. 
\textit{All mutually unbiased bases in dimensions two to five.} Quantum Inf. Comp. 10, 0803 (2010)

\bibitem{BR98}
D. Bru\ss.
\textit{Optimal Eavesdropping in Quantum Cryptography with Six States.}
Phys. Rev. Lett. 81(14): 3018–3021, 1998.
doi: \url{https://doi.org/10.1103/PhysRevLett.81.3018}

\bibitem{grassl2004sic} 
M. Grassl. 
\textit{On SIC-POVMs and MUBs in dimension 6.} 
ArXiv Preprint Quant-ph/0406175. (2004)

\bibitem{hao2019new} N. Hao, Z.H. Li, H.Y. Bai and C.M. Bai.
\textit{A new quantum secret sharing scheme based on mutually unbiased bases. }International Journal of Theoretical Physics, 58, 1249-1261, (2019).

\bibitem{HerzogHibi1993} 
J. Herzog and T. Hibi. 
\textit{Monomial ideals.} 
Graduate Texts in Mathematics, Springer-Verlag London, Ltd., London, 260, (2011)
\url{https://doi.org/10.1007/978-0-85729-106-6}

\bibitem{HerzogHibiOshugi2018} 
J. Herzog, T. Hibi and H. Ohsugi. 
\textit{Binomial ideals.} 
Graduate Texts in Mathematics, Springer, Cham, 279, (2018). \url{https://doi.org/10.1007/978-3-319-95349-6}

\bibitem{ivonovic1981geometrical} 
I. Ivonovic. 
\textit{Geometrical description of quantal state determination.} 
Journal Of Physics A: Mathematical And General, 14, 3241 (1981)

\bibitem{kibler} 
M. R. Kibler and M. Planat. 
\textit{A SU(2) recipe for MUBs.}
Int. J. Mod. Phys, B 20, 1802 (2006)

\bibitem{kolountzakis} 
M. N. Kolountzakis, M. Matolcsi, and M. Weiner. 
\textit{An application of positive definite functions to the problem of MUBs.}
Proc. Amer. Math. Soc., 146, 1143 (2018)

\bibitem{li2019mutually} 
T. Li, L.M. Lai, S.M. Fei and Z.X. Wang. 
\textit{Mutually unbiased measurement based entanglement witnesses.}
International Journal of Theoretical Physics, 58, 3973-3985 (2019).

\bibitem{Matsumura1987} 
H. Matsumura. 
\textit{Commutative Ring Theory.}
Cambridge Studies in Advanced Mathematics, 8, Translated from the Japanese by M. Reid (1986)

\bibitem{paterekdakic} 
T. Paterek, B. Dakic, and C. Brukner. 
\textit{Mutually unbiased bases, orthogonal Latin squares, and hidden variable models.}
Phys. Rev. A 79, 012109 (2009)

\bibitem{paterekpawlowski} 
T. Paterek, M. Pawłowski, M. Grassl and C. Brukner. 
\textit{On the connection between mutually unbiased bases and orthogonal Latin squares.}
Phys. Scr. T140, 014031 (2010)

\bibitem{planat2006survey} 
M. Planat, H. Rosu and S. Perrine. 
\textit{A survey of finite algebraic geometrical structures underlying mutually unbiased quantum measurements.}
Foundations Of Physics, 36, 1662-1680 (2006)

\bibitem{finitepro} 
M. Saniga, M. Planat and H. Rosu. 
\textit{MUBs and finite projective planes.}
J. Opt. B 6, L19 (2004)

\bibitem{schwinger1960unitary} 
J. Schwinger.
\textit{Unitary operator bases.}
Proceedings Of The National Academy Of Sciences, 46, 570-579 (1960)

\bibitem{sun2024applications} 
Y. Sun, M.J. Zhao and P.T. Li.  
\textit{Applications of Geometric Coherence with Respect to Mutually Unbiased Bases.} 
International Journal of Theoretical Physics, 63(10), 264, (2024).

\bibitem{tadej2006concise} 
W. Tadej and K. Zyczkowski. 
\textit{A concise guide to complex Hadamard matrices.}
Open Systems \& Information Dynamics, 13, 133-177 (2006)

\bibitem{tavakoli2015secret} 
A. Tavakoli, I. Herbauts, M. Żukowski and  M. Bourennane.
\textit{Secret sharing with a single d-level quantum system.}
Physical Review A, 92(3), 030302 (2015).

\bibitem{terhal2000bell} 
B. M. Terhal. 
\textit{Bell inequalities and the separability criterion.}
Physics Letters A 271, no. 5-6, 319-326, (2000).

\bibitem{wang2021constructing} K. Wang and Z. J. Zheng. 
\textit{Constructing entanglement witnesses from two mutually unbiased bases.}
International Journal of Theoretical Physics, 60, pp.274-283, (2021).

\bibitem{wootters1989optimal} 
W. Wootters and B. Fields.
\textit{Optimal state-determination by mutually unbiased measurements.}
Annals Of Physics, 191, 363-381 (1989)

\bibitem{zauner2011quantum} 
G. Zauner. 
\textit{Quantum designs: Foundations of a noncommutative design theory.}
International Journal Of Quantum Information, 9, 445-507 (2011)

\end{thebibliography}
\end{document}